\documentclass[12pt]{article}

\usepackage{amsmath,amssymb,amsfonts,amsbsy}
\usepackage{cite}
\usepackage{graphicx}
\usepackage{wrapfig}
\usepackage{epsfig}
\usepackage{bbm} % blackboard bold
\usepackage{bm} % blackboard bold
\usepackage{color}                                                       %
\usepackage{dsfont} % double stroke roman fonts                          %
\usepackage{latexsym} % a few additional special symbols                 %
\usepackage{lscape} % allows single pages to be set landscape            %
\usepackage{mathrsfs} % alternative mathematical "script" font           %
\usepackage{morefloats} % solves excess floating figure problems         %
\usepackage{floatflt} % solves excess floating figure problems           %
\usepackage{slashed} % slash notation for Dirac gamma-matrix products    %
\usepackage{psfrag}

\DeclareFontFamily{OT1}{mygreek}{}%
\DeclareFontShape{OT1}{mygreek}{m}{n}{<->omsegr}{}%
\DeclareFontShape{OT1}{mygreek}{b}{n}{<->omsegrb}{}%
\DeclareFontShape{OT1}{mygreek}{m}{it}{<->omsegri}{}%
\DeclareFontShape{OT1}{mygreek}{bx}{n}{<->sub * mygreek/b/n}{}%
\DeclareFontShape{OT1}{mygreek}{m}{sl}{<->sub * mygreek/m/it}{}%
\DeclareSymbolFont{Greekrm}{OT1}{mygreek}{m}{n}
\DeclareSymbolFont{Greekbf}{OT1}{mygreek}{b}{n}
\DeclareSymbolFont{Greekit}{OT1}{mygreek}{m}{it}
\DeclareMathSymbol{\omegab}{\mathalpha}{Greekbf}{119}
\begin{document}

\addcontentsline{toc}{subsection}{{About spin electromagnetic wave-particle with ring singularity}\\
{\it A.A.~Chernitskii}}

%%%%%%% please do not touch these! %%%%%%
\setcounter{section}{0}
\setcounter{subsection}{0}
\setcounter{equation}{0}
\setcounter{figure}{0}
\setcounter{footnote}{0}
\setcounter{table}{0}

\begin{center}
\textbf{ABOUT SPIN ELECTROMAGNETIC WAVE-PARTICLE\\ WITH RING SINGULARITY}
\vspace{5mm}

\underline{A.A.~Chernitskii}$^{\,1,2}$

\vspace{5mm}

\begin{small}
  (1) \emph{A. Friedmann Laboratory for Theoretical Physics, St.-Petersburg} \\
  (2) \emph{State University of Engineering and Economics,\\ Marata str. 27, St.-Petersburg, Russia, 191002} \\
  \emph{E-mail: AAChernitskii@mail.ru, AAChernitskii@engec.ru}
\end{small}
\end{center}

\vspace{0.0mm} % Don't laugh: it does change the spacing!

\begin{abstract}
An axisymmetric space-localized solution of nonlinear electrodynamics is considered as massive charged particle
with spin and magnetic moment. The appropriate solution for nonlinear electrodynamics with ring
singularity is investigated.
In view of this problem the system of toroidal waves in linear electrodynamics is considered.
The problem with boundary conditions
on the singular ring of the toroidal coordinate system is investigated. The boundary conditions are taken
taking into account the conformity between the toroidal and cylindrical waves on the ring.
In this case the singular ring looks like convolute axis of cylindrical system.
The appropriate system of wave modes are obtained in an integral form with the help of source function.
\end{abstract}

\vspace{7.2mm}

%%%%%%%%%%%%%%%%%%%%%%%%%%%%%%%%%%%%%%%%%%%%%%%
%%%% My definitions
\def\bfgr#1{\pmb{#1}}
\def\p{\partial}
\def\dfrac#1#2{{\displaystyle\frac{#1}{#2}}}
\def\stTD#1#2{\hbox to 0em{\mathsurround=0em $\stackrel{#1}{\makebox[0pt]{} #2}$\hss} \phantom{#2}}\def\stscript#1#2{\hbox to 0em{\mathsurround=0em ${\scriptstyle\stackrel{#1}{\makebox[0pt]{} #2}}$\hss} \phantom{#2}}\def\stscriptscript#1#2{\hbox to 0em{\mathsurround=0em ${\scriptscriptstyle\stackrel{#1}{\makebox[0pt]{} #2}}$\hss} \phantom{#2}}\def\st#1#2{\mathchoice{\stTD{#1}{#2}}{\stTD{#1}{#2}}{\stscript{#1}{#2}}{\stscriptscript{#1}{#2}}}
\def\comb#1#2#3{{\mathsurround 0pt\hbox to 0pt {\hspace*{#3}\raisebox{#2}{${#1}$}\hss}}}
\def\combs#1#2#3{{\mathsurround 0pt\hbox to 0pt {\hspace*{#3}\raisebox{#2}{${\scriptstyle #1}$}\hss}}}
\def\combss#1#2#3{{\mathsurround 0pt\hbox to 0pt {\hspace*{#3}\raisebox{#2}{${\scriptscriptstyle #1}$}\hss}}}
\def\e#1{\mathrm{e}^{#1}}
\def\ii{\mathrm{i}}
\def\veps{\varepsilon}
\def\sech{\operatorname{sech}}
\def\df{\mathrm{d}}
\def\dfs{\mathchoice{\combs{|}{0.45ex}{0.1em}\mathrm{d}}{\combs{|}{0.45ex}{0.1em}\mathrm{d}}{\combss{|}{0.25ex}{0.05em}\mathrm{d}}{}{}}
\def\metr{\mathfrak{m}}
\def\Fem{F}
\def\bFem{\mathbf{F}}
\def\fem{G}
\def\bfem{\mathbf{G}}
\def\bcdot{\mathchoice{\mathbin{\boldsymbol{\comb{\cdot}{0.15ex}{0.37ex}\Diamond}}}{\mathbin{\boldsymbol{\comb{\cdot}{0.05ex}{0.4ex}\Diamond}}}{\mathbin{\boldsymbol{\combs{\cdot}{0.13ex}{0.28ex}\Diamond}}}{}{}}
\def\bwedge{\mathchoice{\mathbin{\combs{\boldsymbol{\backslash}}{0.2ex}{0.1ex}\boldsymbol{\wedge}}}{\mathbin{\combs{\boldsymbol{\backslash}}{0.2ex}{0.1ex}\boldsymbol{\wedge}}}{\mathbin{\combss{\boldsymbol{\backslash}}{0.1ex}{0ex}\boldsymbol{\wedge}}}{}{}}
\def\Vol{\mathchoice{\combs{\square}{0.15ex}{0.2ex} {\rm V}}{\combs{\square}{0.15ex}{0.2ex} {\rm V}}{\combss{\square}{0.12ex}{0.095ex} {\rm V}}{}{}}
\def\bVol{\mathchoice{\combs{\pmb{\boldsymbol{\square}}}{0.15ex}{0.4ex} {\mathbf{V}}}{\combs{\pmb{\boldsymbol{\square}}}{0.15ex}{0.4ex} {\mathbf{V}}}{\combss{\pmb{\boldsymbol{\square}}}{0.1ex}{0.2ex} {\mathbf{V}}}{}{}}
\def\dVol{{\rm d}\hspace{-0.3ex}\Vol}
\def\bdVol{{\rm d}\hspace{-0.3ex}\bVol}
\def\ci{{\rm i}}
\def\sqm{\sqrt{|\metr|}}
\def\Eem{E}
\def\bEem{\mathbf{E}}
\def\Hem{H}
\def\bHem{\mathbf{H}}
\def\Dem{D}
\def\bDem{\mathbf{D}}
\def\Bem{B}
\def\bBem{\mathbf{B}}
\def\je{{\mathsurround 0pt\lower.0ex\hbox{${\scriptscriptstyle e}$}\mspace{-4.5mu}j}}
\def\bje{{\mathsurround 0pt\lower.0ex\hbox{${\scriptscriptstyle \mathbf{e}}$}\mspace{-3.4mu}\mathbf{j}}}
\def\jm{{\mathsurround 0pt\lower.0ex\hbox{${\scriptscriptstyle m}$}\mspace{-7.0mu}j}}
\def\bjm{{\mathsurround 0pt\lower.0ex\hbox{${\scriptscriptstyle \mathbf{m}}$}\mspace{-5.6mu}\mathbf{j}}}
\def\bjem{\mathbf{j}}
\def\baab{\mathbf{b}}
\def\bbaab{{\mathsurround 0pt\mbox{${\bf b}$\hspace*{-1.1ex}${\bf b}$}}}
\def\p{\partial}
\def\bp{\boldsymbol{\p}}
\def\unitc{\mathbf{1}}
\def\him{\imath}
\def\bhim{\boldsymbol{\imath}}
\def\invop#1{{#1^{\!\!-\!1}}}
\def\bHCLor{\boldsymbol{\Lambda}}
\def\bhcLor{\boldsymbol{\lambda}}
\def\bHCLorS{\comb{\boldsymbol{\circ}}{0.15ex}{0.24ex}{\boldsymbol{\Lambda}}}
\def\hcLorS{\comb{\circ}{0.15ex}{0.2ex}{\lambda}}
\def\bhcLorS{\comb{\boldsymbol{\circ}}{0.2ex}{0.2ex}{\boldsymbol{\lambda}}}
\def\commut#1#2{\pmb{\bigl[}\, #1 \,\pmb{|}\, #2\, \pmb{\bigr]}}
\def\hconj#1{\mathchoice{{{}^{\boldsymbol{*}}\mspace{-2mu}#1}}{{{}^{\boldsymbol{*}}\mspace{-2mu}#1}}{{{}^{\boldsymbol{*}}\mspace{-4mu}#1}}{}{}}
\def\bRe{\boldsymbol{\Re}}
\def\bIm{\boldsymbol{\Im}}
\def\bDB{\mathbf{Y}}
\def\bEH{\mathbf{Z}}
\def\bnpc{\bnp^{\circ}}
\def\bnp{\comb{\boldsymbol{\cdot}}{0ex}{0.4ex} {\boldsymbol{\partial}}}
\def\nosum{\mathchoice{\textstyle \comb{\pmb{\bigl/}}{0ex}{1ex} {\sum}}{\comb{\pmb{\bigl/}}{0ex}{0.8ex} {\sum}}{}{}}
\def\prodf#1#2{\mathchoice{\pmb{\Bigl\langle} #1\, \pmb{|}\, #2 \pmb{\Bigr\rangle}}{\pmb{\bigl\langle} #1\, \pmb{|}\, #2 \pmb{\bigr\rangle}}{\pmb{\langle} #1\, \pmb{|}\, #2 \pmb{\rangle}}{}{}}
\def\bShST{\mathchoice{\combs{\rightarrow}{0.2ex}{0.5ex} \boldsymbol{\mathcal{J}}}{\combs{\rightarrow}{0.2ex}{0.5ex} \boldsymbol{\mathcal{J}}}{\combss{\rightarrow}{0.12ex}{0.2ex} \boldsymbol{\mathcal{J}}}{}{}}
\def\bRotS{\mathchoice{\comb{\circ}{0.05ex}{0.78ex} \boldsymbol{\mathcal{J}}}{\comb{\circ}{0.05ex}{0.78ex} \boldsymbol{\mathcal{J}}}{\combs{\circ}{0.05ex}{0.55ex} \boldsymbol{\mathcal{J}}}{}{}}
\def\Fourier#1{{}_{\mathtt{f}}\!\mspace{-1.5mu}#1}
\def\conjop#1{{}^{{\boldsymbol{\diamond}}}\!#1}
\def\cylaf#1#2{\mathchoice{\combs{\|}{0.42ex}{0.63ex}{\mathbf{C}^{#2}_{#1}}}{\combs{\|}{0.42ex}{0.63ex}{\mathbf{C}^{#2}_{#1}}}{\combss{\|}{0.245ex}{0.41ex}{\mathbf{C}^{#2}_{#1}}}{}{}}
\def\wavevp{\mathchoice{{\mathsurround 0pt\mbox{${k}$\hspace*{-1.05ex}${k}$}}}{{\mathsurround 0pt\mbox{${k}$\hspace*{-1.05ex}${k}$}}}{{\mathsurround 0pt\mbox{${\scriptstyle {k}}$\hspace*{-1.05ex}${\scriptstyle {k}}$}}}{{\mathsurround 0pt\mbox{${\scriptstyle {k}}$\hspace*{-1.15ex}${\scriptstyle {k}}$}}}{}}
\def\bwavevp{\mathchoice{{\mathsurround 0pt\mbox{$\mathbf{k}$\hspace*{-1.1ex}$\mathbf{k}$}}}{{\mathsurround 0pt\mbox{$\mathbf{k}$\hspace*{-1.1ex}$\mathbf{k}$}}}{{\mathsurround 0pt\mbox{${\scriptstyle \mathbf{k}}$\hspace*{-1.2ex}${\scriptstyle\mathbf{k}}$}}}{{\mathsurround 0pt\mbox{${\scriptstyle \mathbf{k}}$\hspace*{-1.3ex}${\scriptstyle\mathbf{k}}$}}}{}}
\def\Div{\mathnormal{\mathrm{Div}}}
\def\Curl{\mathnormal{\mathrm{Curl}}}
\def\norma#1{\bigl\|#1\bigr\|}
\def\sphersecf#1#2{\mathchoice{\comb{\circ}{0.24ex}{0.13ex}{\mathbf{S}_{#1}^{#2}}}{\comb{\circ}{0.24ex}{0.13ex}{\mathbf{S}_{#1}^{#2}}}{\combs{\circ}{0.16ex}{0.12ex}{\mathbf{S}_{#1}^{#2}}}{}{}}
\def\zc{\mathchoice{\combs{\circ}{0.15ex}{0.11ex}\mathrm{z}}{\combs{\circ}{0.15ex}{0.11ex}\mathrm{z}}{\combss{\circ}{0.05ex}{0.01ex}\mathrm{z}}{}{}}
\def\spherzonf#1#2#3{\mathchoice{\comb{\circ}{0.24ex}{0.23ex}{\mathbf{Z}_{#1#3}^{#2}}}{\comb{\circ}{0.24ex}{0.24ex}{\mathbf{Z}_{#1#3}^{#2}}}{\combs{\circ}{0.15ex}{0.2ex}{\mathbf{Z}_{#1#3}^{#2}}}{}{}}
\def\spheraf#1#2#3{\mathchoice{\comb{\circ}{0.23ex}{0.5ex}{\mathbf{C}^{#2#3}_{#1}}}{\comb{\circ}{0.23ex}{0.5ex}{\mathbf{C}^{#2#3}_{#1}}}{\combs{\circ}{0.17ex}{0.38ex}{\mathbf{C}^{#2#3}_{#1}}}{}{}}
\def\ClGord#1#2#3#4#5{\mathchoice{\combs{\otimes}{0.42ex}{0.25ex}{\mathrm{C}_{#4#5}^{#1#2#3}}}{\combs{\otimes}{0.42ex}{0.25ex}{\mathrm{C}_{#4#5}^{#1#2#3}}}{\combss{\otimes}{0.23ex}{0.1ex}{\mathrm{C}_{#4#5}^{#1#2#3}}}{}{}}
\def\RadfunC#1#2{\mathchoice{\combs{\boldsymbol{\uparrow}}{0.4ex}{0.55ex}{\mathbf{C}^{#1}_{#2}}}{\combs{\boldsymbol{\uparrow}}{0.4ex}{0.55ex}{\mathbf{C}^{#1}_{#2}}}{\combss{\boldsymbol{\uparrow}}{0.2ex}{0.35ex}{\mathbf{C}^{#1}_{#2}}}{}{}}
\def\RadfunS#1#2{\mathchoice{\combs{\boldsymbol{\uparrow}}{0.4ex}{0.2ex}{\mathbf{S}^{#1}_{#2}}}{\combs{\boldsymbol{\uparrow}}{0.4ex}{0.2ex}{\mathbf{S}^{#1}_{#2}}}{\combss{\boldsymbol{\uparrow}}{0.2ex}{0.1ex}{\mathbf{S}^{#1}_{#2}}}{}{}}
\def\garmcb#1#2#3#4{\mathchoice{{}_#3\!\combs{\combs{\|}{0ex}{0.27ex}{\approx}}{0.35ex}{0.35ex}{\mathbf{C}^{#4}_{#1#2}}}{{}_#3\!\combs{\combs{\|}{0ex}{0.27ex}{\approx}}{0.35ex}{0.35ex}{\mathbf{C}^{#4}_{#1#2}}}{{}_#3\!\combss{\combss{\|}{0.04ex}{0.17ex}{\approx}}{0.2ex}{0.25ex}{\mathbf{C}^{#4}_{#1#2}}}{}{}}
\def\garmsb#1#2#3#4{\mathchoice{{}_#2\!\combs{\comb{\circ}{-0.2ex}{0.06ex}{\approx}}{0.4ex}{0.41ex}{\mathbf{C}^{#3#4}_{#1}}}{{}_#2\!\combs{\comb{\circ}{-0.2ex}{0.06ex}{\approx}}{0.4ex}{0.41ex}{\mathbf{C}^{#3#4}_{#1}}}{{}_#2\!\combss{\combs{\circ}{-0.05ex}{0.13ex}{\approx}}{0.2ex}{0.25ex}{\mathbf{C}^{#3#4}_{#1}}}{}{}}
\def\garmtb#1#2#3#4{\mathchoice{{}_#2\!\combs{\combs{\between}{0.05ex}{0.25ex}{\approx}}{0.35ex}{0.39ex}{\mathbf{C}^{#3#4}_{#1}}}{{}_#2\!\combs{\combs{\between}{0.05ex}{0.25ex}{\approx}}{0.35ex}{0.39ex}{\mathbf{C}^{#3#4}_{#1}}}{{}_#2\!\combss{\combss{\between}{0.04ex}{0.15ex}{\approx}}{0.2ex}{0.25ex}{\mathbf{C}^{#3#4}_{#1}}}{}{}}
\def\toraf#1#2#3{\mathchoice{\comb{\circ}{0.23ex}{0.32ex}{\mathbf{T}^{#2#3}_{#1}}}{\comb{\circ}{0.23ex}{0.33ex}{\mathbf{T}^{#2#3}_{#1}}}{\combs{\circ}{0.17ex}{0.245ex}{\mathbf{T}^{#2#3}_{#1}}}{}{}}
\def\Vols{\mathchoice{\combs{\triangle}{0.3ex}{0.13ex} {\rm V}}{\combs{\triangle}{0.3ex}{0.13ex} {\rm V}}{\combss{\triangle}{0.25ex}{0.03ex} {\rm V}}{}{}}
\def\dVols{{\rm d}\hspace{-0.3ex}\Vols}
\def\x{x}
\def\bx{\mathrm{x}}
\def\xp{x^{\prime}\vphantom{x}}
\def\bxp{\mathrm{x}^{\prime}\vphantom{x}}
\def\delf{{\mathsurround=0ex {\displaystyle\comb{\cdot}{-0.1ex}{0.07em}\delta}}}
\def\Sou{S}
\def\Soub{\mathbf{S}}
\def\dSur{{\rm d}\hspace{-0.3ex}\Sigma}
\def\bdSur{{\bf d}\hspace{-0.3ex}{\bfgr{\Sigma}}}
\def\Sur{\Sigma}
\def\gx{\mathfrak{x}}
\def\bfgx{\boldsymbol{\mathfrak{x}}}
\def\tgx{\tilde{\mathfrak{x}}}
\def\bftgx{\tilde{\boldsymbol{\mathfrak{x}}}}
\def\dsur{{\rm d}\hspace{-0.3ex}{{\sigma}}}
\def\bdsur{{\bf d}\hspace{-0.3ex}{\bfgr{\sigma}}}
\def\bbdsur{{\mathsurround 0pt\mbox{${\bf d}\hspace{-0.3ex}{\bfgr{\sigma}}$\hspace*{-2.3ex}${\bf d}\hspace{-0.3ex}{\bfgr{\sigma}}$}}}
\def\sur{\sigma}
\def\sx{\mathsf{x}}
\def\sxp{\mathsf{x}^\prime\vphantom{\mathsf{x}}}
\def\bbfgx{{\mathsurround 0pt\mbox{$\boldsymbol{\mathfrak{x}}$\hspace*{-1.2ex}$\boldsymbol{\mathfrak{x}}$}}}
\def\bbftgx{\tilde{{\mathsurround 0pt\mbox{$\boldsymbol{\mathfrak{x}}$\hspace*{-1.2ex}$\boldsymbol{\mathfrak{x}}$}}}}
\def\iICpm#1#2#3#4{{}_#1\!\mathcal{I}_{#2,#3,#4}}
\def\tCc#1#2{\tilde{C}_{#1,#2}}
\def\tmp#1{\ \ \breve{\phantom{I}}\!\!\!\!\!\!\!\garmcb{\omega}{#1}{\bumpeq}{l}}
\def\eqdef{\doteqdot}
%%%% End of my definitions
%%%%%%%%%%%%%%%%%%%%%%%%%%%%%%%%%%%%%%%%%%%%%%%

The present work is the part of the work on the construction of the field model for massive charged elementary particle
with spin and magnetic moment as a soliton solution of nonlinear electrodynamics.
This theme was discussed in my articles. See for example
\cite{Chernitskii1999,Chernitskii2004a,Chernitskii2005a,Chernitskii2007a,Chernitskii2008b}

In this approach we have mass and spin of the particle  as three dimensional space integral from the energy and angular momentum densities for electromagnetic field:
\begin{equation}
\mathrm{m}=\int\limits_{V}\mathcal{E}\,dv\;\;,
\qquad \mathrm{s} = \biggl|\int\limits_{V}\mathbf{r}\times\bfgr{\mathcal{P}}\,dv\biggr|\;\;,
\end{equation}
where $\mathcal{E} = \mathcal{E}(\mathbf{D},\mathbf{B})$ is the energy density, $\mathbf{D}$ and $\mathbf{B}$ are electric and magnetic inductions,
$\bfgr{\mathcal{P}}=\dfrac{1}{4\pi}\mathbf{D}\times\mathbf{B}$ is the Poynting vector.
The function $\mathcal{E}(\mathbf{D},\mathbf{B})$ defines the concrete model of nonlinear electrodynamics.

Here we consider the field configuration with ring singularity.

In general case the appropriate soliton solution in own coordinate system has a static part
and quickly-oscillating part.
The static part gives mass, spin, charge, and magnetic moment of the particle.
The oscillating part gives the wave behavior of the particle.

The finding of the appropriate exact solution of nonlinear electrodynamics is the very difficult problem. But we can use approximate methods.
The short report on the investigation of static solution with ring singularity for Born-Infeld nonlinear electrodynamics is contained in my article
\cite{Chernitskii2010a}.

Now we investigate the oscillating part of the soliton solution with ring singularity.
The present work is dedicated to construction the system of undistorted (standing) toroidal waves in linear electrodynamics. The linear waves
can be used in perturbation schemes for finding the soliton solution under consideration with the oscillating part.

It should be noted that the linear problem considered here is not trivial because the variables are not separated in Helmholtz equation for toroidal coordinates.

Here we will use the hypercomplex form for electrodynamics (see
\cite{Chernitskii2002a}).
In this case the Clifford algebra with noncommutative product is used.

The hypercomplex form for representation of electromagnetic bivector
is used:
$\bFem = \bEem  {}+{} \bhim\,\bBem$,
where  $\bhim$ is hyperimaginary unit.

According to my paper
\cite{Chernitskii2003c} we can write
\begin{align}
\bFem (\bx) &= -\frac{1}{4\pi}\int\limits_{\Sur^{\prime}}\Soub (\bx^\prime {}-{} \bx)\,\bdSur^\prime\,\bFem (\bx^\prime)
\;\;,
\label{48185782}
\end{align}
where
$\Soub (\bx)$ is the source function,
$\Sur^\prime$ is
the three-dimensional hypersurface bounding the four-volume,
 and $\bdSur^{\prime} $ is its inside oriented element.

Let us consider that the field $\bFem$ is harmonic wave. Thus we can write
\begin{equation}
\label{67667953}
\bFem = \bFem_\omega\,\e{-\bhim\,\omega\, x^0}
 \;\;.
 \end{equation}
where $\bFem_\omega = \bFem_\omega (\sx)$, $\sx \equiv \{\x^1,\x^2,\x^3\}$.
%, $\omega = 2\pi/T$ is circular frequency.

Let us consider the known toroidal coordinate system in three-dimensional space $\{u,\,v,\,\varphi\}$ with the following transformation formulas to cylindrical coordinates $\{\rho,\,\varphi,\, z\}$:
\begin{equation}
\label{48976015}
\rho = \dfrac{\rho_\circ\,\sinh v}{\cosh v-\cos u}\;\;,\quad
z = \dfrac{\rho_\circ\,\sin u}{\cosh v - \cos u}\;\;,
\end{equation}
where $\rho_\circ$ is the radius of the singular ring,\\
\centerline{ $-\pi< u \leqslant\pi$,\quad
$0\leqslant v <\infty$,\quad $0\leqslant\varphi< 2\pi$.}

We will use the modified toroidal coordinate system $\{\tau,\,\eta,\,\varphi\}$, where
\begin{equation}
\label{37541946}
\begin{array}{ll}
\tau = \sech v\;\;,\quad   &0\leqslant\tau\leqslant 1\;\;,\\
\eta = - u\;\;,\quad   &-\pi <\eta < \pi
  \;\;.
\end{array}
\end{equation}
The coordinate $\tau$ ranges from $0$ to $1$ when the coordinate $v$ ranges from $\infty$ to $0$.

We have the following transformation formulas from the modified toroidal coordinates $\{\tau,\,\eta,\,\varphi\}$ to cylindrical ones
\begin{equation}
\label{48906015}
\rho = \dfrac{\rho_\circ\,\sqrt{1-\tau^2}}{1 - \tau\,\cos\eta}\;,\quad
z = -\dfrac{\rho_\circ\,\sin\eta}{1 - \tau\,\cos\eta}\;,
\end{equation}

 As we can easy obtain
 the behavior of the modified toroidal coordinates near the singular ring ($\tau\to 0$) is the following:
\begin{subequations}
\label{38059361}
\begin{align}
\label{38049525}
  &\bbaab^{\tau} \sim \dfrac{1}{\rho_\circ}\left(-\sin \eta\,\bbaab_{z} +  \cos \eta
   \,\bbaab_{\rho}\right)
 \;,\\[1ex]
\label{38069023}
  &\bbaab^{\eta} \sim -\dfrac{1}{\rho_\circ\,\tau}\left( \cos\eta\,\bbaab_{z} +  \sin\eta\,\bbaab_{\rho}\right)
 \;,
\\[1ex]
&\bbaab^{\varphi} \sim \frac{1}{\rho_\circ}\left(-\sin\varphi\,\bbaab_1 + \cos\varphi\,\bbaab_2\right)
\;,
\\[1ex]
&\metr_{\tau\tau} \sim \rho_\circ^2\;,\quad
\metr_{\eta\eta}\sim \rho_\circ^2\, \tau^2\;,\quad
\metr_{\varphi\varphi}\sim \rho_\circ^2\;
\;,
\\[1ex]
&\dVols \sim \rho_\circ^3\,\tau\, \df \tau\,\df \eta\,\df \varphi
\;,
\end{align}
where $\bbaab^i$ are  the basis bivectors,
$\metr_{ij}$ are the components of metric tensor, $\dVols$  is
the three-dimensional volume element.
\end{subequations}

Let us introduce the following curvilinear coordinates:
\begin{subequations}
\label{39764394}
\begin{align}
   \label{39788689}
 &\breve{\rho}\eqdef\rho_\circ\,\tau\;\;,\quad
  \breve{\varphi} \eqdef \eta \;\;,\quad
  \breve{z}\eqdef \rho_\circ\,\varphi
  \;\;,\\
   \label{39788690}
 &\bbaab^{\breve{\rho}} \eqdef \rho_\circ\,\bbaab^{\tau} \;\;,\quad
 \bbaab^{\breve{\varphi}}\eqdef \bbaab^{\eta}  \;\;,\quad
 \bbaab^{\breve{z}}\eqdef \rho_\circ\,\bbaab^{\varphi}
  \;\;.
\end{align}
\end{subequations}

As we can see in (\ref{39764394}) with (\ref{38059361}) the coordinates $\{\breve{\rho},\,\breve{\varphi},\,\breve{z}\}$ (\ref{39764394}) near the ring looks locally like the cylindrical coordinates.

Thus we will consider that the toroidal wave solutions near ring is close to
the radial-undistorted cylindrical waves propagating along the $z$ axis.
These cylindrical waves obtained in \cite{Chernitskii2003c}
have the following form:
% (see (7.25) in \cite{Chernitskii2005a}):
\begin{equation}
\label{77488096}
\garmcb{\omega}{k_z}{\bumpeq}{m} \e{-\bhim\, \omega\,x^0}
 \;\;,
\end{equation}
where $\garmcb{\omega}{k_z}{\bumpeq}{m} = \garmcb{\omega}{k_z}{\bumpeq}{m} (\rho,\varphi,z )$ are cylindrical bivector eigenfunctions of operator $(-\bhim\,\bnp)$ corresponding undistorted waves (``Bessel beams'', see my paper \cite{Chernitskii2003c})
\begin{equation}
\label{427160581}
 -\bhim\,\bnp\,\garmcb{\omega}{k_z}{\bumpeq}{m} = \omega\,\garmcb{\omega}{k_z}{\bumpeq}{m}
\;\;,\qquad
\bnp\equiv \bbaab^i\,\p_i
\;\;,
\end{equation}
$m$ is the index of the angle cylindrical function, $k_z$ is the wavenumber corresponding to the propagation along the $z$ axis.

The ring play a part of $z$ axis for the toroidal
system. In this case the appropriate wave number $k_\varphi$ (instead of $k_z$ for cylindrical waves) is quantized
because of continuity condition for the field near the ring. Thus we have
\begin{equation}
\label{40197984}
\left.
\begin{array}{l}
2\pi\,\rho_\circ = \left|m\right|\,\lambda_\varphi
\\[1ex]
\left|k_\varphi\right| = \dfrac{2\pi}{\lambda_\varphi}
\end{array}
\right\}
\qquad  \Longrightarrow\qquad
k_\varphi= \frac{m}{\rho_\circ}
 \;\;,
 \end{equation}
where $m$ is integer (as positive as negative values is used), $\lambda_\varphi$ is the wave-length at the ring.

Thus we consider that the toroidal wave near ring has the following form:
\begin{equation}
\label{42595437}
\garmtb{\omega}{\bumpeq}{l}{m}\e{-\bhim\, \omega\,x^0}\sim
\tmp{k_\varphi}
%\ \ \ \ \ \breve{\phantom{I}}\!\!\!\!\!\!\!\garmcb{\omega}{k_\varphi}{\bumpeq}{l}
\e{-\bhim\, \omega\,x^0}
\quad\text{for}\quad \tau\to 0
\;\;,
\end{equation}
where $\garmtb{\omega}{\bumpeq}{l}{m} = \garmtb{\omega}{\bumpeq}{l}{m}(\tau,\eta,\varphi)$ are toroidal bivector eigenfunctions of operator $(-\bhim\,\bnp)$ corresponding undistorted waves,
$\tmp{k_\varphi} = \tmp{k_\varphi} (\breve{\rho}, \breve{\varphi},\breve{z})
= \tmp{\frac{m}{\rho_\circ}} (\rho_\circ\tau, \eta,\rho_\circ\varphi)$ (see (\ref{40197984}) and (\ref{39788689})).

We consider the solutions in the form of
 some kind of standing toroidal waves but which can contain closed traveling waves.
 These closed traveling waves propagate along the ring and around the ring.
 The power flow through any closed surface
 containing the singular ring is absent. We will search these waves in the form
\begin{equation}
\label{77488001}
\garmtb{\omega}{\bumpeq}{l}{m}\e{-\bhim\, \omega\,x^0}
 \;\;.
\end{equation}
%where $\garmtb{\omega}{\bumpeq}{l}{m} = \garmtb{\omega}{\bumpeq}{l}{m}(\tau,\eta,\varphi)$ are
%the toroidal bivector eigenfunctions of operator $(-\bhim\,\bnp)$.

To obtain the functions $\garmtb{\omega}{\bumpeq}{l}{m}$ we use formula (\ref{48185782}) and boundary condition on the toroidal surface near the singular ring according to relation (\ref{42595437}).

Let us consider the toroidal surface $\{\tau = \tau_\circ,\,-\pi< \eta \leqslant\pi,\, 0\leqslant \varphi < 2\pi\}$.
This surface will play the role of two-dimensional part of hypersurface $\Sigma^\prime$ in (\ref{48185782}) such that the appropriate primed coordinates is
$\{\tau^\prime = \tau_\circ,\,-\pi< \eta^\prime \leqslant\pi,\, 0\leqslant \varphi^\prime < 2\pi\}$.

After necessary substitutions we obtain
\begin{align}
\nonumber
\garmtb{\omega}{\bumpeq}{l}{m} (\tau,\eta,\varphi)  &=
%\\
%\nonumber
%&\qquad 
-\frac{1}{8\pi}\,\lim\limits_{\tau_\circ\to 0}\int\limits_{\sur^\prime}
 \left[
 \left(\e{-\bhim\,\omega\, \tgx} + \e{\bhim\,\omega\, \tgx}\right)
 \frac{\omega}{\tgx}\,
 \tmp{\frac{m}{\rho_\circ}}
 %(\rho_\circ\tau_\circ, \eta^\prime,\rho_\circ\varphi^\prime)
 %\garmtb{\omega}{\bumpeq}{l}{m} (\tau^\prime,\eta^\prime,\varphi^\prime)
 \times\bbdsur^{\prime}
\right.
\\
\nonumber
&
\qquad\qquad
+\left(\frac{1}{\tgx^3}
\left(\e{-\bhim\,\omega\, \tgx} + \e{\bhim\,\omega\, \tgx}\right)
%\cos (\omega\,\tgx)
 +\frac{\omega}{\tgx^2}\,
 \bhim\left(\e{-\bhim\,\omega\, \tgx} - \e{\bhim\,\omega\, \tgx}\right)
 %\sin (\omega\,\tgx)
 \right)
\\
&
\qquad\qquad\qquad
\left.\cdot \left(\vphantom{\frac{\omega}{\tgx^2}}\bbftgx\,\left(
\tmp{\frac{m}{\rho_\circ}}
%(\rho_\circ\tau_\circ, \eta^\prime,\rho_\circ\varphi^\prime)
%\bFem_\omega (\sxp)
\cdot\bbdsur^{\prime}\right) +
\bbftgx\times\left(
\tmp{\frac{m}{\rho_\circ}}
%(\rho_\circ\tau_\circ, \eta^\prime,\rho_\circ\varphi^\prime)
%\bFem_\omega (\sxp)
\times\bbdsur^{\prime}\right)\right)\right]
\;,
\label{71975419}
\end{align}
where
$\sur^{\prime}$ is the toroidal surface bounding the singular ring such that $\rho_\circ$ is its big radius and
$\tau_\circ$ is its small radius,
$\bdsur^{\prime}$ is its outside oriented element,
$\tmp{\frac{m}{\rho_\circ}}=\tmp{\frac{m}{\rho_\circ}} (\rho_\circ\tau_\circ, \eta^\prime,\rho_\circ\varphi^\prime)$
is the appropriate cylindrical functions on the surface $\sur^{\prime}$,
$\tgx$ is the distance from the point
$\{\tau,\eta,\varphi\}$ to the point
$\{\tau_\circ,\eta^\prime,\varphi^\prime\}$,
$\bbftgx$ is the appropriate radius bivector.

%$\bdsur^{\prime} \equiv \baab^i\,\dsur^{\prime}_i$, $\bbdsur^{\prime}\equiv \baab_0\,\bdsur^{\prime} = \bbaab^i\,\dsur^{\prime}_i$,
%$\bbftgx \equiv \baab_0\,\bftgx  = \bbaab^i\,\tgx\,\p^\prime_i\,\tgx$ ($\bbftgx = (\xp^i-\x^i)\,\bbaab^i$ for Cartesian coordinates).

Thus here we have the integral representation for toroidal undistorted linear electromagnetic waves. This representation can be sufficient
for the using of these functions.

%\bibliographystyle{abbrv}
%\bibliography{CHERNITSKIIDSPIN2011PROC}

\end{document}